\documentstyle[prb,aps]{revtex} 
\input BoxedEPS.tex
\SetepsfEPSFSpecial
\HideDisplacementBoxes

\begin{document}

\twocolumn[\hsize\textwidth\columnwidth\hsize
          \csname @twocolumnfalse\endcsname
\title{The resistive transition and Meissner effect in carbon 
nanotubes: Evidence for  quasi-one-dimensional superconductivity 
above room temperature} 
\author{Guo-meng Zhao$^{*}$} 
\address{ Department 
of Physics and Astronomy, California State University at Los Angeles, 
Los Angeles, CA 90032, USA}

\maketitle
\widetext
\vspace{0.3cm}

\begin{abstract}

It is well known that copper-based perovskite oxides 
rightly enjoy consensus as high-temperature superconductors on the 
basis of two signatures:  the resistive transition and the Meissner effect. 
We show that the resistive transitions in carbon nanotubes agree 
quantitatively with  the Langer-Ambegaokar-McCumber-Halperin (LAMH) 
theory for quasi-1D superconductors although the superconducting transition 
 temperatures can vary from 0.4 K to 750 K for different samples. We 
 have also identified the 
 Meissner effect in the field parallel to the tube axis up to room temperature 
 for  aligned and physically separated multi-walled nanotubes (MWNTs).  The magnitude of 
 the Meissner effect is in quantitative agreement with the predicted 
 penetration depth from the measured carrier density.
 Furthermore, the 
 bundling of individual MWNTs into closely packed bundles leads to a large 
 enhancement in the diamagnetic susceptibility, which is the hallmark 
 of the Josephson coupling among the tubes in bundles.  These results 
 consistently indicate quasi-1D high-temperature superconductivity in carbon 
 nanotubes. 
~\\
~\\
\end{abstract}
\narrowtext
]
\section{Introduction}

It is well known that copper-based perovskite oxides 
rightly enjoy consensus as high-temperature superconductors on the 
basis of two signatures: the Meissner effect and the  sharp resistive transition 
to the zero resistance state. In 
contrast, these two important signatures are far less obvious in 
a quasi-one-dimensional (quasi-1D) superconducting wire that has a finite number of 
transverse conduction channels and a very thin transverse dimension. 
Due to large superconducting fluctuations, the resistive transition 
in quasi-1D superconductors could be very broad. Because of the finite 
number of transverse channels,  the four-probe resistance never goes 
to zero even though the on-wire resistance approaches zero. Because the 
penetration  depth is far larger than the transverse dimension, the 
Meissner effect becomes very small and less visible. These unique 
features make it difficult to unambiguously identify quasi-1D 
superconductivity in altrathin wires or tubes.

Previously we have provided over twenty pieces of evidence (see two 
review articles \cite{Zhaorev1,Zhaorev2} and references therein) for 
quasi-1D high-temperature superconductivity in individual 
single-walled carbon nanotubes (SWNTs), in SWNT bundles/mats, in 
individual multi-walled nanotubes (MWNTs), and in MWNT mats. 
Here we make quantitative data analyses on the observed 
resistive transitions and magnetic properties in carbon nanotubes. 
We find that the resistive transitions in carbon nanotubes agree 
quantitatively with  the Langer-Ambegaokar-McCumber-Halperin (LAMH) theory 
\cite{Langer} for 
 quasi-1D superconductors although the superconducting transition 
 temperatures vary from 0.4 K to 750 K for different samples. We 
 have also identified the 
 Meissner effect in the field parallel to the tube axis up to room temperature 
 for  aligned MWNTs that are physically separated.  The magnitude of 
 the Meissner effect is in quantitative agreement with the predicted 
 penetration depth from the measured carrier density.
 Furthermore, the diamagnetic susceptibility
 in closely packed MWNT bundles increases by a factor of over 4 at 
 low temperatures compared with that for physically separated tubes.
 This is the hallmark 
 of the Josephson coupling among the tubes in bundles.  These results 
 consistently indicate quasi-1D high-temperature superconductivity in carbon 
 nanotubes. 

\section{Theoretical description for the resistive transition in quasi-1D 
superconductors}

The phenomenon of superconductivity depends on the coherence of the 
phase of the superconducting order parameter.  The phase coherence of 
the superconducting order parameter leads to the zero-resistance state.  
For three-dimensional (3D) bulk systems, the transition to the zero-resistance state occurs 
right below the mean-field superconducting transition temperature 
$T_{c0}$ such that the resistive transition is very sharp and the 
transition width is negligibly small.  In contrast,
the resistive transition in quasi-1D superconductors 
is broad because of large superconducting 
fluctuations.  A quantum theory to describe the resistive transition in 
quasi-1D superconductors was developed by Langer, Ambegaokar, McCumber 
and Halperin (LAMH) \cite{Langer} over 30 years ago.  The theory is 
based on thermally activated phase slips (TAPS), which cause the 
resistance to decrease to zero exponentially. In addition to the thermally 
activated phase slips, there are also quantum phase slips due to 
a finite number of transverse channels \cite{Zaikin}, which prevent altrathin wires 
or tubes from being true superconductors with absolutely zero resistance. 

Very recently, we have shown \cite{Zhao04} that the observed resistive transition of a 
superconducting  
carbon nanotube bundle ($T_{c0}$ = 0.44 K ) is in quantitative agreement with 
the LAMH theory. We have also 
demonstrated that \cite{Zhao04} the  
resistive transition below $T^{*}_{c}$ = 0.90$T_{c0}$ is 
simply proportional to $\exp 
(-\frac{3\beta T^{*}_{c}}{T}(1-\frac{T}{T^{*}_{c}})^{3/2})$, where 
the barrier height has the same form as that predicted by the LAMH 
theory. The quantitative agreement between 
theory and experiment indicates that the LAMH theory can correctly 
describe the resistive transition in the low-$T_{c}$ superconducting carbon 
nanotube bundle.  Then, a natural way to demonstrate high-temperature 
superconductivity in other carbon nanotubes is to see whether  the 
observed resistive transitions agree with the LAMH 
theory in a quantitative way.

For two-probe or four-probe measurements on carbon nanotubes with finite 
transverse channels, the total resistance is $R = R_{0}+R_{tube}$, 
where $R_{tube}$ is  the on-tube 
resistance and $R_{0}$ = $R_{t}$ = $R_{Q}/tN_{ch}$ for four-probe 
measurements, or $R_{0}$ = $R_{Q}/tN_{ch}+R_{c}$ for two probe 
measurements \cite{Sheo}. Here $t$ is the transmission 
coefficient ($t \leq$ 1), $R_{Q}$ = $h/2e^{2}$ = 12.9 k$\Omega$ is 
the resistance quantum, $R_{c}$ is the contact resistance, and   
$R_{t}$ is the tunneling resistance.  Both $R_{c}$ and 
$R_{t}$ should be temperature independent.  For ideal contacts, $R_{c}$ = 0 and 
$t$ = 1,  so $R_{0}$ = $R_{t}$ = 12.9 
k$\Omega$/$N_{ch}$ for a bundle comprising $N_{ch}$ transverse channels. 
For quasi-1D systems,  $N_{ch}$ is always finite such that  both 
four-probe and two-probe resistances  
never go to zero even if the on-tube resistance is zero. Only 
if $N_{ch}$ goes to infinity, as in the bulk 3D systems, $R_{t}$ becomes zero such that 
four probe resistance can go to zero below the superconducting transition 
temperature. Therefore, the non-zero four-probe resistance in 
altrathin tubes does not rule out superconductivity in the tubes.

 According to the LAMH theory, the on-tube resistance is 
 given by \cite{Zhao04}
\begin{equation}\label{Er}
R_{tube} = \frac{m}{T^{1.5}}(1-\frac{T}{T_{c0}})^{9/4}\exp 
[-\frac{3cT_{c0}}{T}(1-\frac{T}{T_{c0}})^{3/2}],
\end{equation}
with
\begin{equation}\label{Em}
m = 2.55T_{c0}(3cT_{c0})^{1/2}[\frac{L}{\xi (0)}], 
\end{equation}
and 
\begin{equation}\label{Ec}
c = 0.34N_{ch}\frac{\xi (0)}{\xi_{BCS}}.
\end{equation}

Here $m$ is in the unit of k$\Omega$K$^{3/2}$, $\xi_{BCS}$  is the 
BCS coherence length, and $\xi (0)$ is the zero-temperature coherence 
length.  In the clean limit, $\xi (0)$ = 0.74$\xi_{BCS}$ and thus 
$c = 0.25N_{ch}$. The estimated region of validity 
for Eq.~\ref{Er}  is below 0.07$R_{N}$ ($R_{N}$ is the normal-state 
resistance) for dirty wires 
where the mean free path $l$ $<$$<$ $\xi_{BCS}$ (Ref.~\cite{MMP}). For 
cleaner wires, the region of validity increases. For example, the 
estimated regions of validity 
are below about 0.17$R_{N}$ and 0.55$R_{N}$ 
for  $l$ = $\xi_{BCS}$ and $l$ = 10$\xi_{BCS}$, respectively.   
Moreover, we have found that \cite{Zhao04} the 
on-tube resistance below $T_{c}^{*}$$\simeq$  0.9$T_{c0}$ can be 
excellently described by 
\begin{equation}\label{Eeon} 
R_{tube} = \alpha\exp 
[-\frac{3\beta T^{*}_{c}}{T}(1-\frac{T}{T^{*}_{c}})^{3/2}]. 
\end{equation}
Here the $\beta$ value is nearly the same as the $c$ value 
\cite{Zhao04}.  The microscopic 
origin of this simple empirical formula is unknown at present.  In the 
following, we will compare these formulas with the measured resistive 
transitions in several different samples with $T_{c0}$'s ranging from 
0.4 K to 750 K.

\section{The resistive transition in a SWNT bundle with $T_{c0}$ = 
0.44 K}

In 2001, Kociak {\em et al.} provided the first 
experimental evidence for superconductivity in single-walled carbon 
nanotube bundles from the electrical transport measurements 
\cite{Kociak}.  Although the 
superconducting transition temperature is low ($<$ 1 K), the resistive behavior 
of the nanotube bundle can serve as a prototype for the resistive 
transition in quasi-1D superconducting wires with a finite number of 
transverse conduction channels.

\begin{figure}[htb]
\ForceWidth{7cm}
	\centerline{\BoxedEPSF{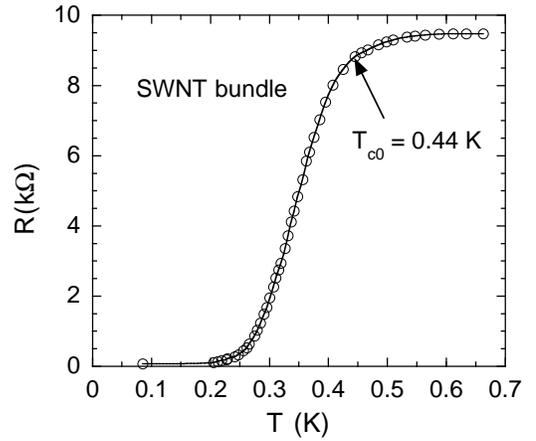}}
	\vspace{0.4cm}
\caption[~]{The temperature dependence of the two-probe resistance for a SWNT bundle that consists of about 
350 tubes.  The data are extracted from Ref.~\cite{Kociak}.}
\end{figure}

Fig.~1 shows the two-probe resistance data for a SWNT bundle that consists of about 
350 tubes \cite{Kociak}.  One can see that the resistance starts to drop below about 
0.5 K, decreases more rapidly below  $T_{c0}$ $\simeq$ 0.44 K and saturates to a value 
of 74 $\Omega$. From the saturated value of $R_{0}$ = 74 $\Omega$, and the 
relation: $R_{0}$= $R_{Q}/tN_{ch}+R_{c}$, one can 
easily find that more than 174 transverse channels are connected to 
the electrodes and participate in electrical transport. This implies that more than 87 metallic-chirality 
superconducting SWNTs take part in electrical transport. Considering 
the fact that one third of tubes should have metallic chiralities and 
become superconducting, we find the total number ($N_{m}$) of the 
superconducting tubes to be 117, implying that $t$ $\geq$ 0.74. The 
value of $N_{m}$ can be also deduced from the measured current 
$I_{c}^{*}$ at which the last resistance jump occurs. The $I_{c}^{*}$ corresponds to 
the critical current for a superconducting wire without disorder and 
with the same number of the transverse channels \cite{Kociak}. For 
metallic chirality superconducting carbon nanotubes, one can readily 
deduce that \cite{Zhao} $I_{c}^{*} = 
7.04k_{B}T_{c0}N_{m}/eR_{Q}$. With $I_{c}^{*}$ = 2.4 $\mu$A 
(Ref.~\cite{Kociak}) and  $T_{c0}$ = 
0.44 K, we have $N_{m}$ = 116, in remarkably good agreement with the 
value deduced above.

In Fig.~2, we fit the resistance data below 
0.88$T_{c0}$ by 
\begin{equation}\label{Ee} 
R = R_{0}+\alpha\exp 
[-\frac{3\beta T^{*}_{c}}{T}(1-\frac{T}{T^{*}_{c}})^{3/2}]. 
\end{equation}
Here the first term is the sum of the tunneling and contact 
resistances and equal to 74 $\Omega$, and the second term is the on-tube 
resistance (see Eq.~\ref{Eeon}).  We can see that the fit is excellent 
with the fitting parameters $\beta$ = 2.99$\pm$0.05 and $T^{*}_{c}$ = 0.394$\pm$0.002 K. Reducing or increasing the temperature region for the fit tends to 
worsen the fit quality. Therefore, the on-tube resistance 
goes to zero exponentially below $T^{*}_{c}$ = 0.9$T_{c0}$. In fact, 
the on-tube resistance may never go to zero if we consider quantum 
phase slips due to the finite number of transverse channels. 
Nevertheless, the on-tube resistance well below $T_{c0}$ should be negligibly small if 
$N_{ch}$ is not so small. 

\begin{figure}[htb]
\ForceWidth{7cm}
	\centerline{\BoxedEPSF{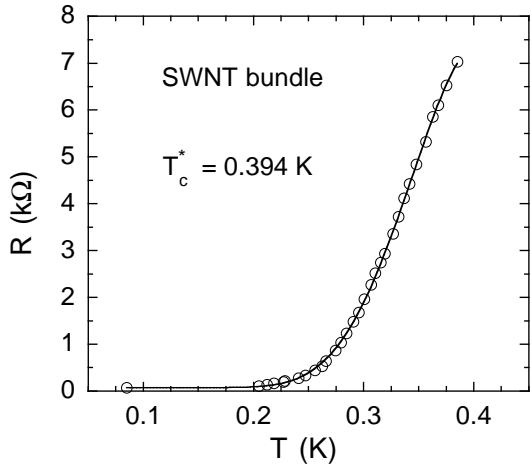}}
	\vspace{0.3cm}
\caption[~]{The temperature dependence of the two-probe resistance for 
a SWNT bundle below 0.88$T_{c0}$. 
The solid line is the curve best fitted by Eq.~\ref{Ee} 
with $\beta$ = 2.99$\pm$0.05 and $T^{*}_{c}$ = 0.394$\pm$0.002 K. 
Reducing or increasing the temperature region for the fit tends to 
worsen the fit quality. It is striking that the 
on-tube resistance below 0.88$T_{c0}$ decreases exponentially to 
zero.}
\end{figure}

In Fig.~3, we fit the resistance data 
below 0.06$R_{N}$ by the following equation. 
\begin{equation}\label{MH} 
R = 74+\frac{m}{T^{1.5}}(1-\frac{T}{T_{c0}})^{9/4}\exp 
[-\frac{3cT_{c0}}{T}(1-\frac{T}{T_{c0}})^{3/2}].
\end{equation}
Here the second term is the on-tube resistance which is the same as 
Eq.~\ref{Er} predicted by the LAMH theory.  We fit the resistance data 
in this temperature region  because $l$ $<$$<$ $\xi_{BCS}$ is well satisfied 
in this SWNT bundle, as seen below. One can see that the 
fitting is very good with the fitting parameters: $m$ = 26.6$\pm$4.7 k$\Omega 
K^{1.5}$ and $c$ = 3.08$\pm$0.13. It is remarkable that the value of 
$c$ is nearly the same as the value of $\beta$ (2.99$\pm$0.05) deduced 
above from a simple exponential fit (Eq.~\ref{Ee}).  

\begin{figure}[htb]
\ForceWidth{7cm}
	\centerline{\BoxedEPSF{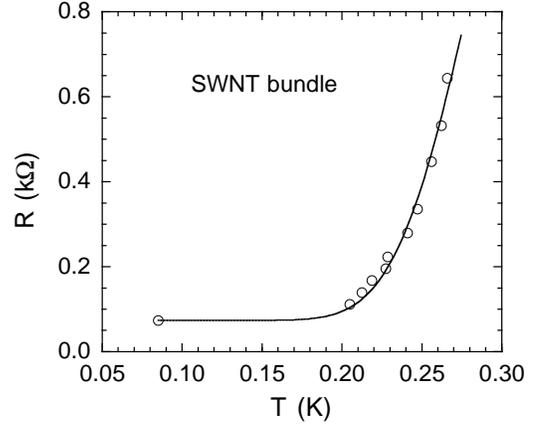}}
	\vspace{0.2cm}
\caption[~]{The temperature dependence of the two-probe resistance for a SWNT 
bundle below 0.06$R_{N}$. 
The solid line is the curve fitted to the data below 0.06$R_{N}$ by 
Eq.~\ref{MH} predicted by the LAMH theory. The estimated region of validity 
for the  LAMH theory is below 0.07$R_{N}$ for dirty wires where 
$l$ $<$$<$ $\xi_{BCS}$ (Ref.~\cite{MMP}). The condition of $l$ $<$$<$ $\xi_{BCS}$ is well satisfied in the SWNT bundle (see text). }
\end{figure}

From the fitting 
parameters $c$ and $m$, and the measured normal-state resistance, we have 
deduced \cite{Zhao04} $l$ = 46 \AA, $\xi (0)$ = 850 \AA, and $\xi_{BCS}$ = 21739 \AA.  Using these values, we can calculate 
$c$ = 3.11 from Eq.~\ref{Ec} and  the critical current \cite{Zhao04} $I_{c}$ = 62.4 
nA at 0.1 K.  The calculated value of $c$ is nearly the same as that (3.08) 
deduced from the fitting, and the calculated value of $I_{c}$ is 
in quantitative agreement with the measured value (62 nA) at 
0.1 K (Ref.~\cite{Kociak}).  Moreover, we can 
estimate the Fermi velocity $v_{F}$ from the deduced value 
of $\xi_{BCS}$ and the formula $\xi_{BCS} = 0.18\hbar v_{F}/k_{B}T_{c0}$. 
With $\xi_{BCS}$ = 21739 
\AA~ and $T_{c0}$ = 0.44 K, we get  $\hbar v_{F}$ = 4.6 eV\AA. Then 
we estimate $\gamma_{\circ}$ = 2.16 eV from the formula $\hbar v_{F} = 
1.5a_{C-C}\gamma_{\circ}$ (Ref.~\cite{Mintmire}). This value is very close to the value  
(2.4 eV) estimated from the first-principle calculation \cite{Mint92}. Thus, the 
resistance data of the SWNT bundle agrees with the LAMH theory in a quantitative way.

\section{The resistive transition in a single MWNT  with $T_{c0}$ = 
262 K}

In 1996, Ebbesen {\em et al.} made the first four-probe resistance 
measurements on individual  multi-walled carbon nanotubes \cite{Eb}.  Four  80-nm-wide 
tungsten 
leads were patterned by ion-induced deposition of tungsten from 
W(CO)$_{6}$ carrier gas.  This technique makes it possible for 
electrodes to connect multi-shells of the tubes \cite{Eb1}. It is interesting that 
the electrical properties vary significantly  from samples to 
samples. Some tubes show abrupt jumps in resistivity when the 
temperature increases. In some other tubes, the resistance at room temperature 
is very small  (e.g., 200 $\Omega$) but a metal-insulator transition 
occurs below about 200 K.  We will show that the resistive behavior in the former 
case is in quantitative agreement with that expected for 
quasi-1D superconductivity. The resistive 
behavior in the latter case may be explained by a superconductor-to-insulator  
transition  in dirty quasi-1D systems.

\begin{figure}[htb]
\ForceWidth{7cm}
\centerline{\BoxedEPSF{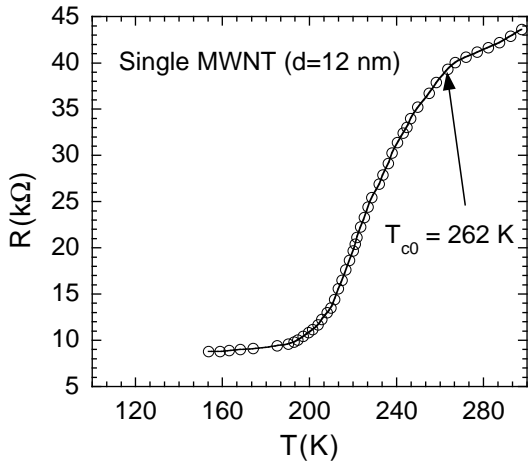}}
	\vspace{0.6cm}
\caption [~]{The 
temperature dependence of the resistance for a single MWNT 
with $d$ $\simeq$ 12 nm. The data are extracted from Ref.~\cite{Eb}. }
\label{fig4}
\end{figure}

Fig.~4 shows the four-probe resistance data for a single MWNT with a 
diameter of 12$\pm$2 nm.  The inner contact distance is $L$ = 5000 \AA. One can see 
that the resistance drops more rapidly 
below  about 262 K and saturates to a value 
of about 8.80 k$\Omega$ below 160 K. The resistive behavior of this single tube 
is similar to the resistive transition for a quasi-1D 
superconductor  with $T_{c0}$ $\simeq$ 262 K. The finite resistance far below the 
superconducting transition  
temperature is due to a finite number of transverse 
channels, which is estimated to be about 54  for this tube (see below). 
Using  $R_{0}$ = 8.80 k$\Omega$, $N_{ch}$ = 54, and 
the relation: $R_{0}$= $R_{Q}/tN_{ch}$, we estimate the average $t$ 
for each channel to be about 0.013.  The small value of $t$ suggests 
that the electrical contacts to the tube are rather poor.  Assuming 
a negligible on-tube resistance below 160 K, we 
estimate that the on-tube resistance in the 
normal state (at 300 K) is about 34 k$\Omega$, leading to $R_{N}/L$= 68 
k$\Omega$/$\mu$m.

In Fig.~5 we fit the resistance data below $T$ = 233 K = 0.89$T_{c0}$ by 
Eq.~\ref{Ee} with a fixed $R_{0}$ = 8.80 k$\Omega$.  One can see that the 
fitting is excellent with the fitting parameters: 
$\beta$ = 11.71$\pm$0.12,  $\alpha$=19.0$\pm$0.1 k$\Omega$, and $T_{c}^{*}$ = 234 K. It is 
interesting that $T_{c}^{*}$ = 0.89$T_{c0}$, the same as that for the 
SWNT bundle with $T_{c0}$ = 0.44 K.

\begin{figure}[htb]
\ForceWidth{7cm}
\centerline{\BoxedEPSF{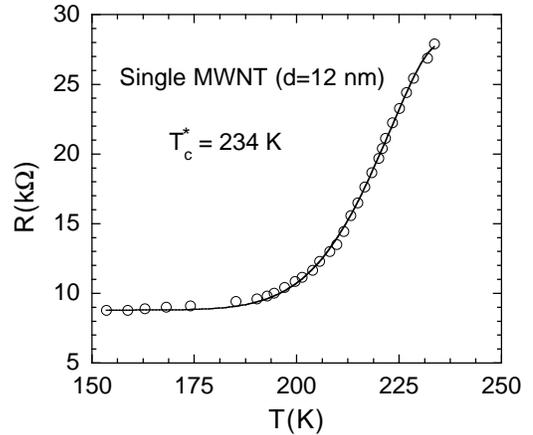}}
	\vspace{0.2cm}
\caption [~]{The 
temperature dependence of the resistance for the 12-nm MWNT 
below 233 K = 0.89$T_{c0}$. 
The solid line is the curve best fitted by Eq.~\ref{Ee} 
with the fitting parameters: $\beta$ = 11.71$\pm$0.12, 
$\alpha$=19.0$\pm$0.1 k$\Omega$, and $T_{c}^{*}$ = 234 K, and with a 
fixed $R_{0}$ = 8.80 k$\Omega$.
Reducing or increasing the temperature region for the fit tends to 
worsen the fit quality. }
\label{fig5}
\end{figure}

In Fig.~6 we fit the resistance data below about 0.15$R_{N}$ by 
\begin{equation}\label{MH2} 
R = 8.80+\frac{m}{T^{1.5}}(1-\frac{T}{T_{c0}})^{9/4}\exp 
[-\frac{3cT_{c0}}{T}(1-\frac{T}{T_{c0}})^{3/2}].
\end{equation}
We fit the resistance data 
in this region  because $l$ $\simeq$ $\xi_{BCS}$,  as seen below. We 
can see that the fitting is excellent between 190 K and 210 K.  There is 
a small deviation between 160 and 190 K, which may arise 
from quantum phase slips. The fitting parameters are
$m$ = (1.64$\pm$0.14)$\times$10$^{7}$ k$\Omega K^{1.5}$ and $c$ = 
10.27$\pm$0.23. It is striking that the values of $\beta$ and $c$ are 
also very close, similar to the case of the SWNT bundle with $T_{c0}$ = 0.44 K.

\begin{figure}[htb]
\ForceWidth{7cm}
\centerline{\BoxedEPSF{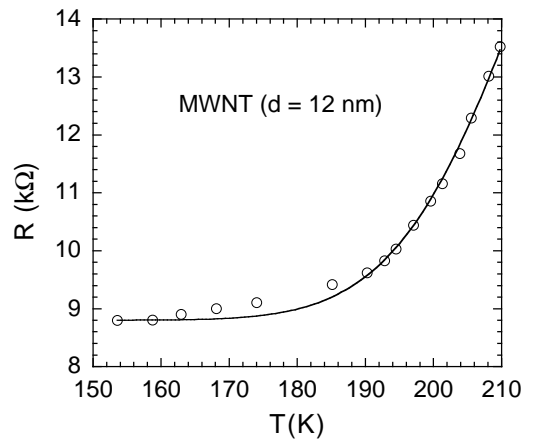}}
	\vspace{0.6cm}
\caption [~]{The 
temperature dependence of the resistance for the 12-nm MWNT 
below about 0.15$R_{N}$. 
The solid line is the curve best fitted by Eq.~\ref{MH2} 
with the fitting parameters: $m$ = (1.64$\pm$0.14)$\times$10$^{7}$ k$\Omega K^{1.5}$ and $c$ = 
10.27$\pm$0.23. }
\label{fig6}
\end{figure}

From the values of $m$, $c$, and $T_{c0}$, we can evaluate the 
zero-temperature coherence length $\xi (0)$ using Eq.~\ref{Em}. 
Substituting $m$ = 
1.64$\times$10$^{7}$ k$\Omega 
K^{1.5}$, $c$ = 10.27, $T_{c0}$ = 262 K, and $L$ = 5000 \AA~ into 
Eq.~\ref{Em}, we obtain $\xi (0)$ = 18.3 \AA.  From the measured 
$R_{N}/L$ = 68 k$\Omega$/$\mu$m and the relation 
$R_{N}/L$ = $R_{Q}/N_{ch}l$ (Ref.~\cite{Bend}), we can calculate 
$N_{ch}l$ = 1897  \AA.  Substituting $c$ =10.27  into  Eq.~\ref{Ec}, we 
get $N_{ch}$ = 30.2$\xi_{BCS}/\xi (0)$.  By solving the three equations: $N_{ch}l$ = 1897  
\AA, $N_{ch}$ = 30.2$\xi_{BCS}/\xi (0)$, and $\xi (0) = 0.74\xi_{BCS}\sqrt{\chi (0.882\xi_{BCS}/l)}$, we 
finally obtain  $l$ = 35 \AA, $N_{ch}$ = 54, and $\xi_{BCS}$ = 33.0 
\AA. Here the Gorkov function $\chi (x)$ is defined as \cite{MMP}
\begin{equation}\label{Echi} 
\chi (x)=\sum_{n=0}^{\infty}\frac{0.95}{(1+2n)^{2}(1+2n+x)}.
\end{equation}
 The fact that $l$ $\simeq$ $\xi_{BCS}$ indicates that the LAMH theory 
(Eq.~\ref{MH2}) is valid only below 0.17$R_{N}$ (Ref.~\cite{MMP}). This justifies the 
region of the data we select to fit. 

We can also evaluate the Fermi velocity $v_{F}$ from the deduced value 
of $\xi_{BCS}$ and the formula $\xi_{BCS} = 0.18\hbar v_{F}/k_{B}T_{c0}$. 
With $\xi_{BCS}$ = 33 \AA~ and $T_{c0}$ = 262 K, we get  $\hbar v_{F}$ = 
4.14  eV\AA.  Using  the value of  $\gamma_{\circ}$ = 2.4 eV estimated 
from the first principle calculation \cite{Mint92} and the relation $\hbar v_{F} = 
1.5a_{C-C}\gamma_{\circ}$ (Ref.~\cite{Mintmire}) for  the first subband of metallic chirality tubes, we  
obtain  $\hbar v_{F}$ = 5.1 eV\AA. It is remarkable that the value of $\hbar 
v_{F}$ deduced from the LAMH theory is about 20$\%$ lower than the 
value expected for the first subband of metallic chirality tubes. Such a small 
discrepancy may be explained by 
a fact that the Fermi level of some outer shells is crossing the second or 
higher subband where \cite{Saito} the Fermi velocity is  smaller 
than the one for the first subband. This quantitative 
agreement provides compelling evidence 
for high-temperature superconductivity at 262 K in this 12-nm MWNT. 

Now we show that the deduced  $N_{ch}$ = 54  is also reasonable. 
For the MWNT with $d$ =12 nm, the total number of shells can be 
estimated to be about 17  using the fact that the intershell distance is 
0.34 nm. This implies that the average number of the conducting 
channels per shell is about 3.2.  This is possible when the Fermi level
of some outer shells crosses their second subband (see above), which have 6 
channels for a 
metallic chirality shell and 4 channels for a semiconducting chirality 
shell \cite{Mintmire}. 

In fact, we can estimate the lower limit of the average number of conducting 
channels per shell for an 18-nm MWNT from the measured room-temperature 
four-probe resistance, which  is 200 $\Omega$ (Ref.~\cite{Eb}).  Using the relation 
$R =  R_{Q}/tN_{ch} + R_{tube}$ $\geq$ $R_{Q}/N_{ch}$, and from $R$ = 200 $\Omega$, we 
get $N_{ch}$ $\geq$ 64. The total number of shells 
for the 18-nm MWNT should be about 26. This implies that the average 
number of conducting 
channels per shell for the 18-nm MWNT is larger than 2.5, in 
agreement with that (3.2) deduced independently for the 12-nm MWNT.

One of the puzzling features of the resistive behavior for this 18-nm 
MWNT is the metal-insulator transition below about 200 K 
(Ref.~\cite{Eb}).  We can 
attribute this to a superconductor-insulator transition in 
dirty quasi-1D systems. It is shown that when the thermal 
length is larger than the localization length below a 
temperature $T_{loc}$ in quasi-1D systems, the Anderson  localization sets in 
and the ground state becomes insulating charge density wave (CDW) 
\cite{Orignac}.  If we make a 
heterojunction between this  insulating MWNT and other metal, we should 
expect a rectification effect below $T_{loc}$. The rectification effect 
will  disappear above $T_{loc}$, in contrast to the 
conventional rectification effect which vanishes at a temperature far 
higher than $E_{g}/k_{B}$, where $E_{g}$ is the gap of a semiconductor. 
Kim {\em et al.} made a heterojunction between an insulating MWNT and a 
``conducting'' MWNT \cite{Kim}. The diameters of both tubes are 30 nm. The 
single-particle tunneling spectrum \cite{Kim} 
indicates that $E_{g}$ for the insulating tube is 
about 150 meV. It is very unusual that \cite{Kim} the rectification 
effect disappears above 192 K. This cannot be explained by the conventional 
model because 192 K  is far below $E_{g}/k_{B}$ = 1740 K.  Furthermore,  according to a formula $E_{g} = 
2a_{C-C}\gamma_{\circ}/d$ (Ref.~\cite{Mintmire}), the 
predicted $E_{g}$ 
for a semiconducting chirality tube with $d$ = 30 nm should be about 22 
meV, which is about one order 
of magnitudes smaller than the observed gap for the insulating tube.  
On the other hand,  the tunneling spectrum for another ``conducting" tube \cite{Kim} is 
consistent 
with $E_{g}$ $\simeq$ 20 meV,  suggesting that the ``conducting'' tube  
has a semiconducting  chirality.  In order to consistently explain 
these novel behaviors, we must assume that the insulating tube has a 
metallic chirality and undergoes a transition from the superconducting to the insulating CDW 
ground state below $T_{loc}$ $\simeq$ 192 K.  It is naturally expected 
that the observed metal-insulator 
transition below about 200 K  in the 18-nm MWNT \cite{Eb} 
should have the same origin as the 30-nm MWNT.  Then, the half gap
$E_{g}/2$ $\simeq$ 75 meV for the insulating tube should be related to 
the minimum single-particle excitation gap in  the CDW ground state. This also 
implies that the minimum superconducting gap would 
be about 75 meV if this 30-nm  MWNT were clean enough to avoid the 
Anderson localization. The value of the superconducting gap suggests 
$T_{c0}$ $>$ $E_{g}/3.52k_{B}$ $\simeq$  500 K, in agreement with the 
resistive transition in a MWNT mat (see below).

\section{The resistive transition in a SWNT mat  with $T_{c0}$ = 
710 K}

Single-walled carbon nanotubes,  prepared by metal-catalysed
laser ablation of graphite, form closely-packed crystalline bundles. 
The bulk  samples, or mats consist of entangled  bundles that are 
contacted each other and oriented randomly \cite{Nature}.  If close-packed crystalline 
bundles are superconductors, the bulk samples should behave like granular  
superconductors. Depending on the Josephson coupling strength between 
the superconducting ``grains'', the ground state could be metallic, 
insulating, or superconducting \cite{Merchant}. It is interesting that the contact 
barrier resistance 
of a granular superconductor follows a rather unusual exponential 
temperature dependence in a certain temperature range \cite{Merchant}, that is, 
$R _{b}(T) =R_{b}(0)\exp (BT)$, where $B$ could be positive, negative, 
or zero.  This temperature dependence of the barrier resistance was 
also suggested \cite{Baum} for the intertube barrier resistance in MWNTs. 
The barrier resistance extrapolated to zero temperature is finite 
even if the $T$-dependence of the resistance behaves like an 
insulator. This unique resistive behavior makes a clear distinction 
from that for conventional semiconductors where the resistance at 
zero temperature goes to infinity.

Fig.~7 shows the temperature dependence of the resistivity for a SWNT mat. The data are extracted from 
Ref.~\cite{Nature}.  Below 200 K the resistivity is nearly 
 temperature independent while above 200 K the resistivity increases 
 suddenly and starts to flatten out above 550 K. This behavior is 
 similar to that for a granular superconductor. Below we will show 
 that this is indeed the case.
 
 \begin{figure}[htb]
\ForceWidth{7cm}
\centerline{\BoxedEPSF{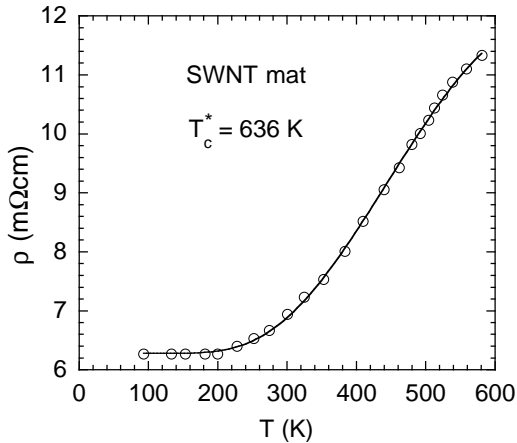}}
	\vspace{0.6cm}
	\caption [~]{The temperature dependence of the resistivity for a 
SWNT mat. The data are extracted from Ref.~\cite{Nature}. The solid 
line is the curve fitted by Eq.~\ref{Eer} with the 
fitting parameters:  $\rho_{0}$= 6.277$\pm$0.018 m$\Omega$cm, 
$\beta$ = 0.90$\pm$0.04, and $T_{c}^{*}$ = 636.3 K.}
\label{4Q}
\end{figure}
 
 We fit the resistivity data by 
\begin{equation}\label{Eer} 
\rho = \rho_{0} (T) +\alpha\exp 
[-\frac{3\beta T^{*}_{c}}{T}(1-\frac{T}{T^{*}_{c}})^{3/2}]. 
\end{equation}
Here the first term is the intertube contact barrier resistivity, which could 
be temperature dependent, and the second term is the on-tube resistivity 
that follows a simple exponential form same as Eq.~\ref{Eeon}. In the 
present case, the intertube barrier resistivity appears to be independent 
of temperature in the temperature region we are interested in, i.e., 
$B$ $\simeq$ 0. It is striking that the data can be well fitted by Eq.~\ref{Eer} with the 
fitting parameters:  $\rho_{0}$= 6.277$\pm$0.018 m$\Omega$cm, 
$\beta$ = 0.90$\pm$0.04, and $T_{c}^{*}$ = 636.3 K. Using the 
relation $T_{c}^{*}$ = 0.895$T_{c0}$ deduced empirically above, we 
obtain $T_{c0}$ = 710 K.  It is remarkable that the 
$T_{c0}$  value obtained from the resistivity data is very close to 
that (665 K) inferred  from the Raman data for a similarly prepared SWNT 
mat \cite{Zhaorev1,Zhaorev2}.

By extrapolation of the data shown in Fig.~7 to $T$ = 710 K, we estimate 
the normal-state on-tube resistivity  at 710 K to be $\rho^{exp}_{N}$$\simeq$ 
1.1$\rho^{exp}_{N}(T_{c0}) $= 6270 $\mu\Omega$cm.  Since about 
one-third of tubes have metallic chiralities and the mat consists of crystalline 
bundles that are 
oriented randomly, the 
intrinsic normal-state on-tube resistivity  $\rho^{i}_{N}$ of the  superconducting  
tubes should be much smaller 
than $\rho^{exp}_{N}$= 6270 $\mu\Omega$cm, that is, 
$\rho^{i}_{N}$ = $\rho^{exp}_{N}/f$, where $f$ is the 
reduction factor that should be close to 3(1/0.33) = 9.  The 
intrinsic mean free path $l$ is related to $\rho^{i}_{N}$ by
\begin{equation}\label{El}
l=\frac{R_{Q}}{2}\frac{A_{\circ}}{\rho_{i}},
\end{equation}
were $A_{\circ}$ is the area of single tube, which is equal to  
1.54$\times$10$^{-18}$ m$^{2}$ for $d$ =1.4 nm.  If we take $f$ = 9, 
we get $l$ = 14 \AA~ at T = 710 K from Eq.~\ref{El}. If the resistivity jump above 
200 K were due to inelastic scattering, the inelastic mean free path 
would be about 14 \AA~ at about 700 K. This is inconsistent with any 
electrical transport mechanism for SWNTs.

Using $\hbar v_{F}$ = 4.5 eV\AA ~and $T_{c0}$ = 710 K, we find that
the BCS coherence length along the 
tube-axis direction $\xi_{BCS}$ = 13.2 \AA.  The zero-temperature 
coherence length along the tube-axis direction $\xi (0)$ should be 
smaller than the clean-limit value: 0.74$\xi_{BCS}$= 10 \AA.  Then the coherence 
length perpendicular to the tube axis should be order of 1 \AA. Such a  short 
coherence length implies that only those superconducting tubes 
that are adjacent to each other can have enough Josephson coupling 
to form a superconducting bundle. A simulation \cite{Stahl} indicates 
the average number of the 
metallic-chirality tubes that are adjacent to each other is about 2. 
This implies that the average number of the 
metallic-chirality tubes comprising 
a superconducting bundle is also about 2, and that there are 
a number of independent superconducting bundles within a physical bundle.

If we assume that the normal-state resistivity is linearly 
proportional to $T$ above 200  K, the average mean-free path between 
200 K and  580 K  should be about 26 \AA, significantly larger than 
$\xi_{BCS}$. Then we estimate $\xi (0)$ $\simeq$ 8 \AA. Using 
Eq.~\ref{Ec} and $c$ $\simeq$ $\beta$ = 0.9, we obtain $N_{ch}$ 
$\simeq$ = 4.5. This implies that, on average, about two metallic chirality tubes 
are adjacent to each other and form a superconducting bundle, in 
quantitative agreement with the simulation \cite{Stahl}. 

\section{The resistive transition in a MWNT mat  with $T_{c0}$ = 
752 K}

Multi-walled carbon nanotubes are prepared by arc discharge of 
graphite.  A multi-walled carbon nanotube is packed in such a way that each shell is 
concentric with each other.  If each shell has phase-incoherent 
superconductivity,   MWNTs are almost optimally  packed to 
maximize the Josephson coupling and phase coherence. Individual  MWNTs 
can be closely packed into bundles. The bulk  samples, or mats are 
made of entangled  bundles that are 
contacted each other and oriented randomly.  The bulk samples should 
also behave like granular  superconductors.

Fig.~8 shows the temperature dependence of the resistance for a MWNT 
mat.  It is interesting that the resistance decreases monotonically 
with increasing temperature below about 570 K. Above 570 K, the 
resistance tends to turn up. The resistance between 300 K and 450 K 
can be excellently described by 17.3$\exp (-T/618.3)$ $\Omega$. This 
temperature dependence is expected for an intertube barrier 
resistance \cite{Baum}.  This implies that the on-tube resistance 
between 300 K and 
450 K is negligible, in agreement with several independent 
experiments which consistently show a negligible on-tube resistance 
at room temperature in many individual MWNTs \cite{Frank,Pablo,Heer,Urb}. The 
observed finite and very small on-tube resistances in the individual MWNTs \cite{Heer} are 
consistent with quasi-1D room-temperature superconductivity with 
finite quantum phase slips.

\begin{figure}[htb]
\ForceWidth{7cm}
\centerline{\BoxedEPSF{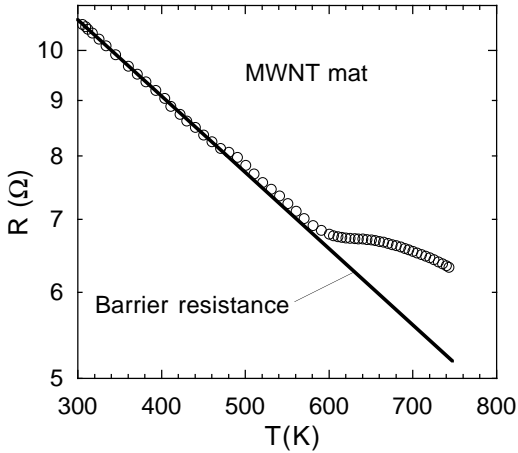}}
	\vspace{0.6cm}
\caption [~]{The 
temperature dependence of the resistance for a MWNT mat. The data are 
the same as  those in  Ref.~\cite{Zhao01} but less data points are 
plotted for charity. The resistance between 300 K and 450 K 
can be excellently described by 17.3$\exp (-T/618.3)$ $\Omega$, which 
represents the intertube barrier resistance \cite{Baum}. }
\label{fig8}
\end{figure}
If we assume that this temperature dependence for 
the intertube barrier resistance remains valid up to 750 K, we then 
obtain the on-tube resistance by subtracting the barrier resistance 
from the total resistance. The resultant on-tube resistance below 665 
K is shown  in Fig.~9. The solid line is the fitted curve by 
Eq.~\ref{Ee} by excluding the data points between 450 K and 600 K. 
The shoulder feature between 450 K and 600 K may be caused by quantum 
phase slips or by bad electrical contacts.
The fitting parameters are $\beta$ = 11.34 
and $T_{c}^{*}$= 669 K. Using the 
empirical relation $T_{c}^{*}$ = 0.89$T_{c0}$, we obtain  $T_{c0}$=752 
K.

\begin{figure}[htb]
\ForceWidth{7cm}
\centerline{\BoxedEPSF{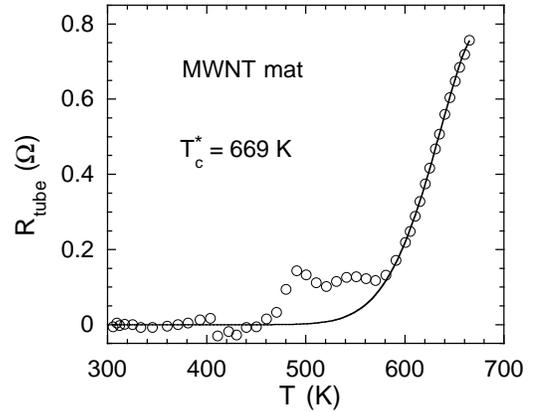}}
	\vspace{0.6cm}
\caption [~]{The on-tube resistance by subtracting the 
barrier resistance 
from the total resistance of the  MWNT mat. The solid line is the fitted curve by 
Eq.~\ref{Ee} by excluding the data points between 450 K and 600 K. 
The fitting parameters are $\beta$ = 11.34 
and $T_{c}^{*}$= 669 K. }
\label{fig9}
\end{figure}

It is interesting that the value of $\beta$ = 11.34  for this MWNT 
mat is slightly smaller than that (11.71) for the 12-nm MWNT.  This implies that, on 
average,  the 
total number of transverse channels for each superconducting bundle 
(which may contain one or more MWNTs near $T_{c0}$)  is 
comparable with that for the single 12-nm MWNT.

\begin{figure}[htb]
\ForceWidth{7cm}
\centerline{\BoxedEPSF{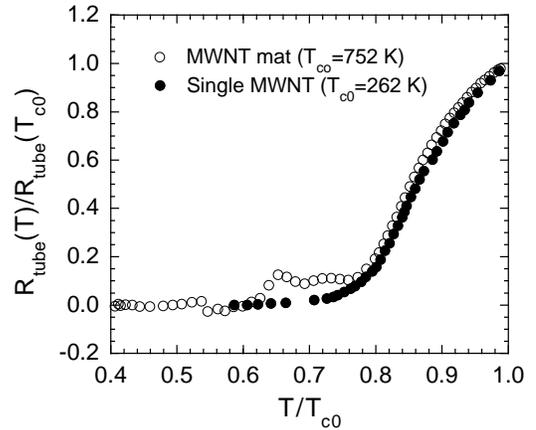}}
	\vspace{0.6cm}
\caption [~]{The normalized on-tube resistance versus 
$T/T_{c0}$ for the 12-nm MWNT ($T_{c0}$ = 262 K) and for the 
MWNT mat ($T_{c0}$ = 752 K).  The on-tube resistive 
transitions for the two systems are nearly identical although they 
have different $T_{c0}$'s. }
\label{fig10}
\end{figure}

In Fig.~10, we plot the normalized on-tube resistance versus 
$T/T_{c0}$ for the 12-nm MWNT ($T_{c0}$ = 262 K) and for the 
MWNT mat ($T_{c0}$ = 752 K). It is remarkable that the on-tube resistive 
transitions for the two systems are nearly identical although they 
have different $T_{c0}$'s and the electrical measurements were done by 
independent groups.  This agreement also suggests that both data sets 
are reliable and that the procedure to extract the on-tube resistance for 
the MWNT mat is justified.

\section{Magnetic properties of MWNTs}

From the quantitative analyses of the electrical transport data in 
several carbon nanotubes, we can clearly see that the superconducting 
transition temperatures can vary from 0.4 K to 750 K for different samples. 
We believe that the $T_{c0}$  variation may be associated with the 
differences in the doping level, 
the chirality  and diameter of tubes, and in disorders. A similar 
$T_{c0}$ 
variation is seen in the graphite-sulfur  system; $T_{c0}$ varies from 
9 K to 250 K (Ref.~\cite{Kop}). In order to unambiguously show that 
high-temperature superconductivity  
in carbon nanotubes is real, one needs to provide magnetic evidence such as 
the Meissner effect. 
However, the Meissner effect may be less visible because the diameters of 
the tubes may be much smaller than the magnetic penetration depth.  
Further, the orbital diamagnetic susceptibility in the magnetic field 
perpendicular to the graphite sheet is large, making it difficult to 
separate the Meissner effect from the large orbital diamagnetic susceptibility. 
Fortunately,  the 
orbital diamagnetic susceptibility of carbon nanotubes in the magnetic field 
parallel to the tube axis is predicted to be very small at room temperature 
\cite{Lu}. 
This makes it possible to extract the Meissner effect from the 
measured susceptibility in the parallel field.
\begin{figure}[htb]
\ForceWidth{7cm}
\centerline{\BoxedEPSF{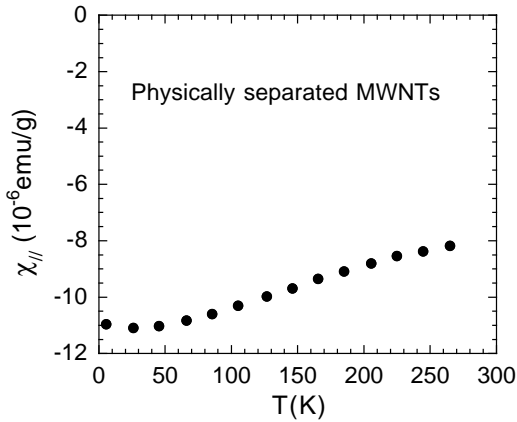}}
	\vspace{0.6cm}
\caption [~]{The  temperature dependence of the susceptibility for 
physically separated and aligned MWNTs in a magnetic field parallel to 
the tube axis. The data are extracted from Ref.~\cite{Chau}. }
\label{fig11}
\end{figure}
Fig.~11 shows the  temperature dependence of the susceptibility for 
physically separated and aligned MWNTs in a magnetic field parallel to 
the tube axis. The diameters of the tubes are 10$\pm$5 nm, and the 
lengths are on the order of 1 $\mu$m. It is apparent that the diamagnetic 
susceptibility  is  
significant up to 265 K.  Because 
the orbital diamagnetic susceptibility in the parallel field is 
negligible at room temperature \cite{Lu}, the observed diamagnetic susceptibility 
at 265 K should mainly contribute from the Meissner effect due to 
superconductivity. Thus, the Meissner effect at 265 K is 
about -0.8$\times 10^{-5}$ emu/g, which is significant.  This result 
clearly indicates that the superconducting 
transition temperature should be higher than 300 K.  

\begin{figure}[htb]
\ForceWidth{7cm}
\centerline{\BoxedEPSF{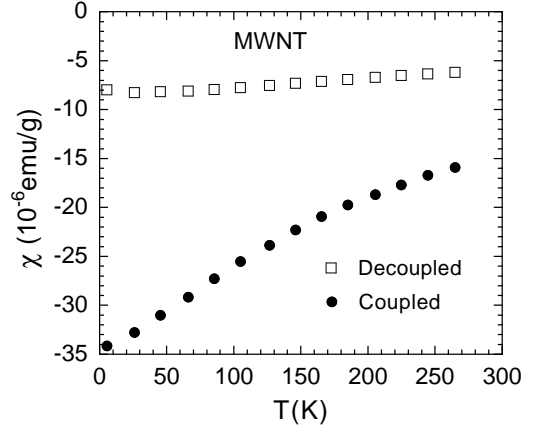}}
	\vspace{0.6cm}
\caption [~]{The  temperature dependence of the angle-averaged susceptibility for 
physically separated MWNTs (open squares) and for physically coupled MWNT 
mat (solid circles). The physically 
coupled mat  sample is not processed so that the tubes are closely 
packed into bundles.  The data are extracted from Ref.~\cite{Chau}. }
\label{fig12}
\end{figure}

For superconducting tubes of radius 
$r$ in the magnetic field parallel to the tube axis, the diamagnetic 
susceptibility due to the Meissner effect is given by
\begin{equation}\label{dia}
 \chi^{S}_{\parallel} (T) = 
-\frac{\bar{r^{2}}}{32\pi\lambda_{\theta}^{2}(T)}.
\end{equation}
Here $\bar{r^{2}}$ is the average value of $r^{2}$, and 
$\lambda_{\theta} (T)$ is the penetration depth when carriers move 
along the circumferential direction.  The above equation is valid only if 
$\lambda _{\theta} (0)$ is larger than the maximum radius of tubes, 
which should be the case for carbon nanotubes. Eq.~\ref{dia} indicates that  
the Meissner effect is inversely proportional to $1/ 
\lambda_{\theta}^{2}(T)$.  Assuming an isotropic gap and 
taking $T_{c0}$ = 752 K, we find that $1/ 
\lambda_{\theta}^{2}(T)$ and thus $\chi^{S}_{\parallel} (T) $ are nearly independent of 
temperature below 265 K. Then we have $\chi^{S}_{\parallel} (0) $
= -0.8$\times 10^{-5}$ emu/g. 
If we assume that the radii of tubes are equally distributed from 0 to 
100~\AA, we find $\bar{r}$ = 50 \AA~and $\bar{r}^{2}$= 3333~\AA$^{2}$.  With the weight 
density of 2.17 g/cm$^{3}$ (Ref.~\cite{Qian}) and $\chi_{\parallel} (0) = 
-0.8\times 10^{-5}$ emu/g, we calculate $\lambda_{\theta} (0)$ 
$\simeq$ 1380 \AA.  The value of the penetration depth corresponds to 
$n/m_{\theta}^{*}$ = $1.48\times 10^{21}/$cm$^{3}m_{e}$, where $n$ is 
the carrier density, $m_{\theta}^{*}$ is the effective mass of 
carriers along the circumferential direction.  If we take 
$m_{\theta}^{*}$ = 0.012 $m_{e}$, typical for graphites \cite{Bayot}, we 
estimate $n$ = 
1.78$\times 
10^{19}/$cm$^{3}$, in good agreement with the Hall effect 
measurement \cite{Baum} which gives $n$ = 1.6$\times 
10^{19}/$cm$^{3}$. It is worthy of noting that the Hall coefficient in 
the physically separated MWNTs does not go to zero below $T_{c0}$. This is because the on-tube resistance does not exactly go to zero due 
to quantum phase slips and because the magnetic field is almost 
penetrated into the whole volume of the tubes.

From Eq.~\ref{dia}, we can see that $\chi^{S}_{\parallel} (T) $ will 
increase linearly with increasing $\bar{r^{2}}$ or cross-sectional 
area.  For Josephson 
coupled MWNT bundles in unprocessed mats, the effective $\bar{r^{2}}$ is 
larger than that for physically separated tubes. As the temperature 
decreases, the Josephson coupling strength increases so that  
the effective $\bar{r^{2}}$  and $\chi^{S}_{\parallel} (T) $ also increases.  This can 
naturally explain why the  diamagnetic 
susceptibility for physically coupled  MWNTs is  larger than that for 
physically separated MWNTs and why the enhancement in the 
diamagnetic susceptibility increases significantly  with decreasing 
temperature (see Fig.~12).  At the lowest temperature, the enhancement factor is 
about 4.3.   Without superconductivity in these MWNTs, it is very difficult to account for 
such a large enhancement in the diamagnetic susceptibility upon 
bundling of  the tubes. 

 \section{Conclusion}

In summary, we have made quantitative data analyses on the observed 
resistive transitions and magnetic properties in carbon nanotubes. 
We show that the resistive transitions in various carbon nanotube 
samples with $T_{c0}$ varying from 0.4 K to 750 K all agree 
with  the theoretical predictions for 
 quasi-1D superconductors in a quantitative way.  We 
 have also identified the 
 Meissner effect in the field parallel to the tube axis up to room temperature 
 for  aligned MWNTs that are physically separated.  The magnitude of 
 the Meissner effect is in quantitative agreement with the predicted 
 penetration depth from the measured carrier density.
 Furthermore, the diamagnetic susceptibility
 in closely packed MWNT bundles increases by a factor of over 4 at 
 low temperatures compared with that for physically separated tubes.
 This is the hallmark 
 of the Josephson coupling among the tubes in bundles.  These results 
 consistently indicate quasi-1D high-temperature superconductivity in carbon 
 nanotubes. 

~\\
~\\
* Correspondence should be addressed to gzhao2@calstatela.edu.

\end{document}